\documentclass[prl,twocolumn,amsmath,superscriptaddress,amssymb,showpacs]{revtex4}

\usepackage{color} 
\usepackage{graphicx}
\usepackage{epstopdf}
\usepackage{bm}
\usepackage{gensymb}

\begin{document}

\DeclareGraphicsExtensions{.eps, .png, .pdf}
\bibliographystyle{prsty}

\def\V2O3{V$_2$O$_3$}
\def\EF{$E_\textrm{F}$}
\def\invA{\AA$^{-1}$}
\def\etal{$et\ al.$}
\def\a1g{$a_{1g}$}
\def\epg{$e_g^{\pi}$}

\title {Fermi Surface of Metallic V$_2$O$_3$ from Angle-Resolved Photoemission:  \\
Mid-level Filling of $e_g^{\pi}$ Bands}

\author{I. Lo Vecchio}
\affiliation{Materials Sciences Division, Lawrence Berkeley National Laboratory, Berkeley, California 94720, USA}

\author{J. D. Denlinger}
\email[]{jddenlinger@lbl.gov}
\affiliation{Advanced Light Source, Lawrence Berkeley National Laboratory, Berkeley, California 94720, USA}

\author{O. Krupin}
\affiliation{Advanced Light Source, Lawrence Berkeley National Laboratory, Berkeley, California 94720, USA}

\author{B. J. Kim}
\affiliation{Max-Planck-Institut fur Festk\"orperforschung, Heisenbergstr. 1, D-70569 Stuttgart, Germany}

\author{P. A. Metcalf}
\affiliation{School of Materials Engineering, Purdue University, West Lafayette, Indiana 47907, USA}

\author{S. Lupi}
\affiliation{CNR-IOM and Dipartimento di Fisica, Universit\`a di Roma ``Sapienza",  I-00185 Rome, Italy}

\author{J. W. Allen}
\affiliation{Randall Laboratory of Physics, University of Michigan, Ann Arbor, Michigan 48109, USA}

\author{A. Lanzara}
\email[]{alanzara@lbl.gov}
\affiliation{Materials Sciences Division, Lawrence Berkeley National Laboratory, Berkeley, California 94720, USA}
\affiliation{Department of Physics, University of California Berkeley, Berkeley, California 94720, USA}

\date{\today}

\begin{abstract}
\noindent Using angle resolved photoemission spectroscopy (ARPES) we report the first band dispersions and distinct features of the bulk Fermi surface (FS) in the paramagnetic metallic phase of the prototypical metal-insulator transition material \V2O3.  Along the $c$-axis we observe both an electron pocket and a triangular hole-like FS topology, showing that both V 3$d$ \a1g\ and \epg\ states contribute to the FS. These results challenge the existing correlation-enhanced crystal field splitting theoretical explanation for the transition mechanism and pave the way for the solution of this mystery.
\end{abstract}
\maketitle

Since its seminal report in 1969 \cite{mcwhan-69,mcwhan-70,mcwhan-73}, the metal-insulator transition (MIT) in the alloy system (V$_{1-x}$Cr$_x$)$_2$O$_3$ has stood as a mystery for many decades.
For $x$=0 and with decreasing temperature ($T$) there is a transition from a paramagnetic metal (PM) to an antiferromagnetic insulator (AFI).  With increasing $x$ the AFI phase persists but the PM phase gives way to a paramagnetic insulator (PI) along an ($x,T$) line terminating at higher $T$ in a critical point. The initial identification of the latter transition as the long sought experimental example of the Mott MIT \cite{mott-68} inherent in the one-band Hubbard model was quickly challenged \cite{goodenough-71} on the grounds that the complexity of the actual multi-orbital electronic structure must be essential for the transition.  This complexity consists of four V$^{3+}$ ions per rhombohedral unit cell with each ion having two 3$d$ electrons to distribute in the two lowest energy trigonal crystal field split 3$d$ states, an orbital singlet \a1g\ and an orbital doublet \epg. Scenarios for reducing this complexity back to a one-band model \cite{castellani-78} were eventually abandoned after X-ray absorption spectroscopy (XAS) showed that the V$^{3+}$ ions are in a Hund's rule S=1 state and that both the \epg\ and \a1g\ states are always occupied, albeit with different occupation ratios $e/a$(=\epg:\a1g) in the three phases \cite{park-00}.  
 
The advent of dynamic mean field theory (DMFT) \cite{georges-96} combined \cite{held-07} with band structure from density functional theory (DFT), supported by bulk sensitive angle integrated photoemission spectra for (V$_{1-x}$Cr$_x$)$_2$O$_3$  \cite{mo-03, mo-06}, gave the first real hope that the mystery could be solved within a realistic multi-orbital calculation.
Indeed a series of DFT+ DMFT studies in the first decade of this century \cite{held-01,laad-03,poteryaev-07} gradually coalesced around a narrative in which the MIT is enabled by a strong many-body enhancement of the trigonal crystal field splitting and thus the orbital polarization of the quasi-particle (QP) bands based on the  \epg\ and \a1g\ states.  In 2007 the claim in Ref. \cite{poteryaev-07}  ``to have demystified the nature of the metal-insulator transition in \V2O3'' seemed well justified by the consensus. 
The study suggested that in the PM phase the Fermi surface (FS) is formed entirely from an \a1g\ QP band, while all the \epg\ QP bands lie entirely below the Fermi energy, \EF, a scenario consistent with electron counting only by virtue of QP weights sufficiently reduced from 1 and the concomitant presence of the lower and upper Hubbard bands below and above \EF, respectively \cite{QPweights}. This strongly polarized and strongly correlated situation brings the PM phase to the very brink of the MIT and the opening of a complete \epg\ - \a1g\ gap. 
Since then however, technical advances in DMFT aimed at achieving full charge self-consistency find much reduced orbital polarizations \cite{grieger-14, deng-14, leonov-15}, implying that what had seemed to be a finished narrative might actually be far from the true implication of DMFT for this problem. 

All this detailed DMFT effort for \V2O3\ has been conducted without any guidance whatsoever from experimental $k$-resolved information on the QP band structure, i.e. no angle resolved photoemission spectroscopy (ARPES). In this Letter we report the first such ARPES data for the PM phase. 
We find that both \a1g\ and \epg\ contribute to the FS, with \EF\ cutting roughly through the center of the \epg\ manifold, i.e. a much weaker orbital polarization. Our results reveal the PM phase electronic structure and suggest that the correlation-enhanced crystal field splitting plays a less important role in driving the MIT than was previously thought. We find some agreement with recent fully charge self-consistent DFT+DMFT calculations, thus demonstrating that this could be the way to solve the \V2O3\ mystery.

\begin{figure}[t]
\leavevmode
\centering
\begin{center}
\includegraphics[width=8.5cm]{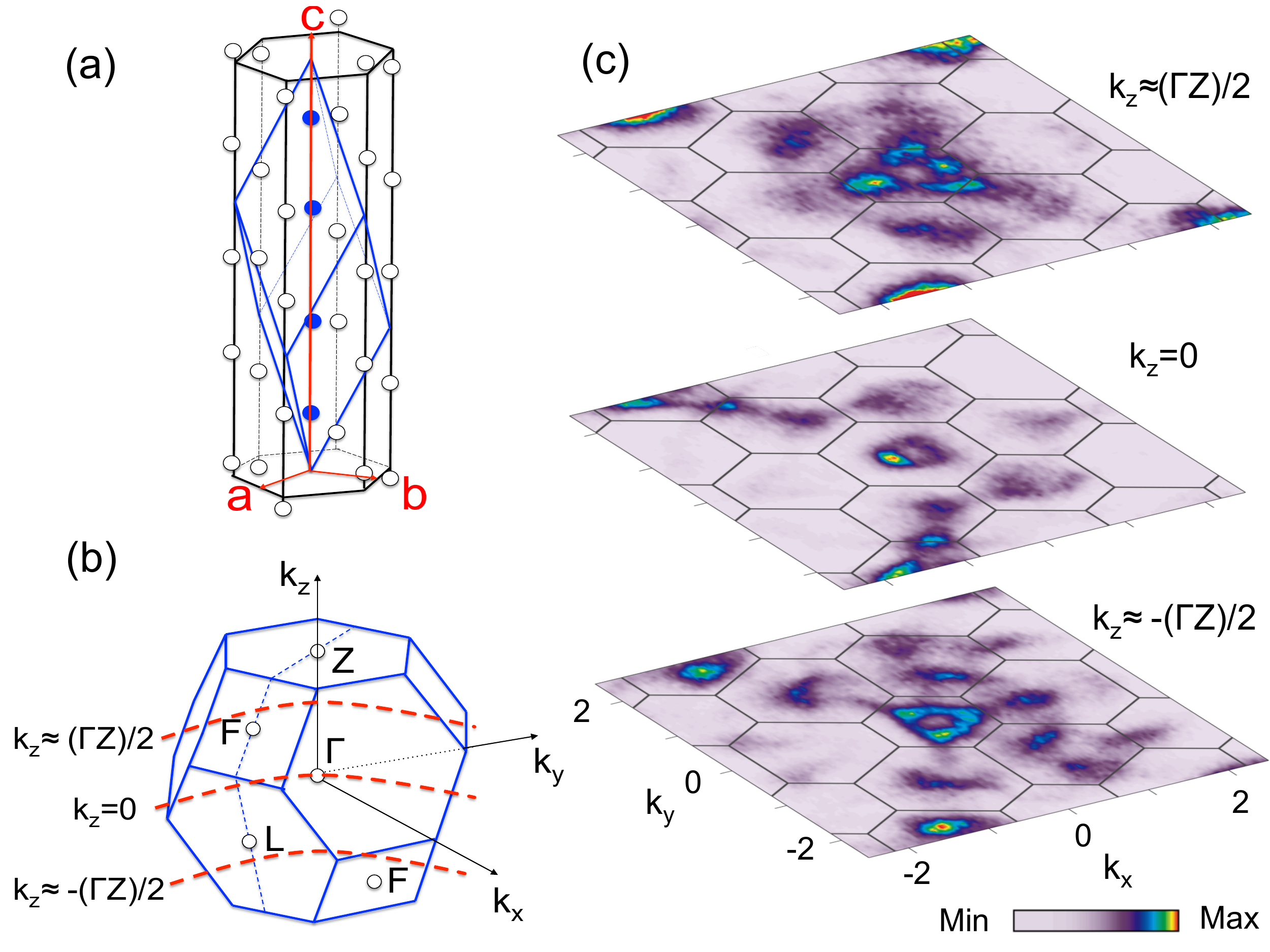}  
\end{center}
\caption{
(a) Rhombohedral primitive cell containing two V$_2$O$_3$ formula units (four blue V atoms), surrounded by the non-primitive hexagonal unit cell. The red hexagonal unit vectors measure a=4.94 \AA\ and  c=13.99 \AA. (b) Brillouin zone for the rhombohedral unit cell of V$_2$O$_3$ paramagnetic metallic phase. (c) Electronic structure of a V$_2$O$_3$ (0001) polished single crystal along normal emission for T=200 K. Wide Fermi surface slices in the plane at the upper half BZ height (360 eV); in the $\Gamma$ plane (320 eV), $k_z$= 9.456 \invA\ but labeled as $k_z$= 0 for simplicity; and in the plane at the lower half BZ height (280 eV), as indicated by dashed curves in (b).
}
\label{Fig1}
\end{figure}

Flat (0001) and (1000) surfaces of V$_2$O$_3$ suitable for ARPES measurements were prepared by annealing polished Laue-oriented single crystals to 750$^\circ$C in a 10$^{-6}$ Torr oxygen partial pressure resulting in 1$\times$1 surface order measured by low energy electron diffraction. Also, room temperature blade cleaving of millimeter sized (0001)-oriented single crystals revealed similar data as for polished single crystals, but with less reproducibility of spectral clarity due to spatial variations \cite{fujiwara-15}. The ARPES measurements were performed using the 2009 configuration of Beamline 7.0 of the Advanced Light Source
utilizing photon energies from 80 eV to 900 eV and a beam spot-size of 50$\times$50 $\mu$m. Data were acquired using a 
Scienta R4000 hemispherical electron analyzer with the sample placed in a vacuum better than 1$\times$ 10$^{-10}$ Torr. The transport MIT transition temperature for these samples is T$_{MIT}$=165 K with a hysteresis of $<$10 K.  Thus the measurements were taken at T=200 K in order to be fully in the PM phase \cite{lupi-10}.

Fig. 1(a) shows the vanadium atom only crystal structure of  V$_2$O$_3$ with two unit cell representations. The primitive rhombohedral unit cell contains two formula units (four solid blue circles along the $c$-axis) while the non-primitive hexagonal unit cell contains six formula units and a $c$-axis notation of (0001). Fig. 1(b) shows the corresponding reciprocal space rhombohedral Brillouin zone with high symmetry point labeling.
Fig. 1(c) presents the first ARPES result of three constant energy maps ($k_x$-$k_y$) of the Fermi-edge intensity acquired for different values of $k_z$ perpendicular to the (0001) surface (accesed by varying the photon energy between 280 and 360 eV). 
The $k_z$ locations \cite{kz} of the maps relative to the bulk BZ correspond to $k_z$=9.456 \invA\ $\Gamma$-plane,  and half-way between $\Gamma$ and Z ($k_z$=$\pm\Gamma$Z/2), as shown schematically with dashed arc lines in Fig. 1(b). 
In the $\Gamma$-plane, a nearly six-fold intensity pattern is observed in the first and second BZs, whereas at $k_z$=$\pm\Gamma$Z/2 an overall three-fold intensity pattern is evident with a strong intensity triangular-like contour in the first BZ. The directional pointing of the triangular FS intensity is observed to reverse in the maps above and below the $\Gamma$-plane. This reversal is consistent with the symmetry of the bulk BZ and provides strong evidence for the bulk origin of these states. 

\begin{figure}[t]
\leavevmode
\centering
\begin{center}
\includegraphics[width=8.6cm]{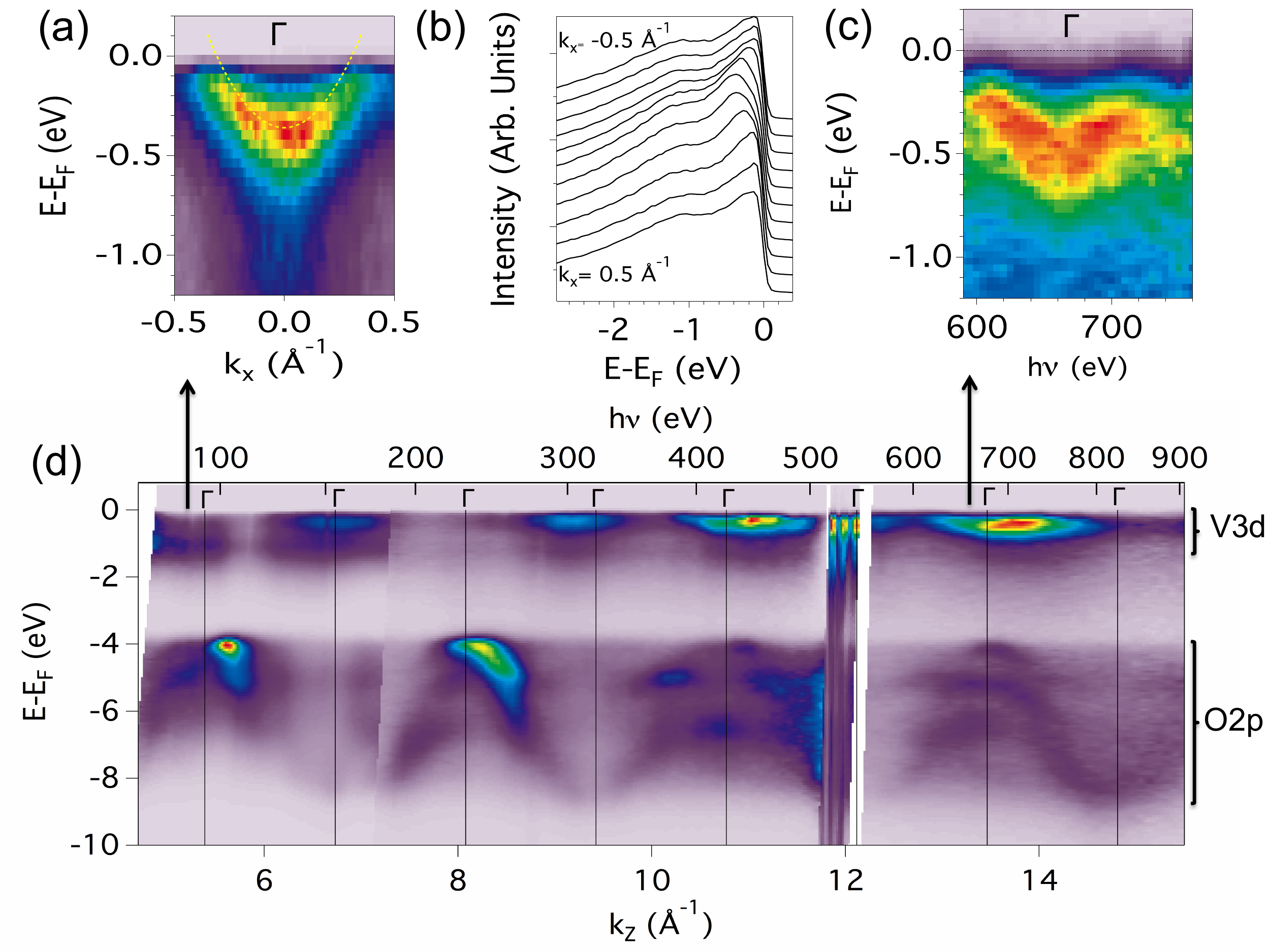}  
\end{center}
\caption{Angle dependent $\Gamma$ point electron pocket dispersion (a) image and (b) stack plot measured at h$\nu$=88 eV for a (1000) surface. (c) Photon dependent $\Gamma$ point electron pocket measured at h$\nu$=670 eV for the (0001) surface. (d) V$_2$O$_3$ (0001) photon energy scan along normal emission, highlighting V 3\textit{d} intensities down to -2 eV and O 2\textit{p} bands dispersing between -4 eV and -8 eV. Vertical lines indicate locations of $\Gamma$ points along the $c$-axis.
 }
\label{Fig2}
\end{figure}

Fig. 2 shows the electronic structure for the (0001) and (1000) surfaces.  Panel (d) shows the energy dispersion as a function of the out-of-plane momentum $k_z$ for normal emission with respect to the (0001) surface (i.e. along the $c$ axis), over multiple Brillouin zones, by varying the photon energy over a 800 eV wide energy range \cite{kz}.  The wide binding energy scale shows both the dispersions of the O 2\textit{p} bands between 4 and 8 eV and the V 3\textit{d} states below 1 eV. The $\Gamma$ points (Brillouin zone center) for each BZ are indicated by a vertical line. The photon dependence is interrupted between 510-560 eV due to strong variations from the V L- and oxygen K-absorption edges. Both the O 2\textit{p} band dispersion and the V 3\textit{d} states near \EF\ show a spectral intensity variation with half unit-cell periodicity. These indicate that the half unit-cell is the primary $c$-axis periodicity of the potential felt by oxygen orbitals.  At more surface sensitive lower photon energies the O 2\textit{p} band dispersion resembles a single 4 eV wide sinusoidal band whereas at higher photon energy the increasing bulk-sensitivity results in greater partitioning of the O 2\textit{p} states into sub-bands \cite{grieger-14}.

Fig. 2(c) shows an expanded zoom of the V 3\textit{d} dispersion around 670 eV, showing an electron pocket centered at $\Gamma$. The reduced visibility of the vanadium states at lower photon energy may be related to the existence of a structural relaxed and/or vanadyl surface termination \cite{feiten-15} resulting in a non-metallic half-unit cell surface and a surface dead layer\cite{borghi-09}. Nevertheless, electron pockets can be observed at low photon energy $\Gamma$-points for the orthogonal (1000) polished surface with the $c$-axis in plane and the unit cell 
 $b$-vector perpendicular to the surface. We conjecture that no such insulating dead layer exists for this (1000) surface. Fig 2(a) shows the $\Gamma$-point electron pocket for this surface measured at 88 eV which allows finer quantification of the electron dispersion to be 0.4 eV deep and 0.75 \AA$^{-1}$ wide. The Fig. 2(b) stack plot of spectra illustrates the relative amplitude of the dispersing component of electron pocket states at 88 eV relative to a broad non-dispersing component of the QP peak and relative to the incoherent band between 0.5 eV and 1.5 eV. The presence of a non-dispersive component in the metallic QP energy region is in part why ARPES of V$_2$O$_3$ has been such a challenging task. Also it produces the visual artifact of the dispersion peak (in color intensity images) not reaching the Fermi level, whereas the metallic dispersion to \EF\ is confirmed by momentum distribution curves cuts of the data set(s).

\begin{figure}[b]
\leavevmode
\centering
\begin{center}
\includegraphics[width=7.4cm]{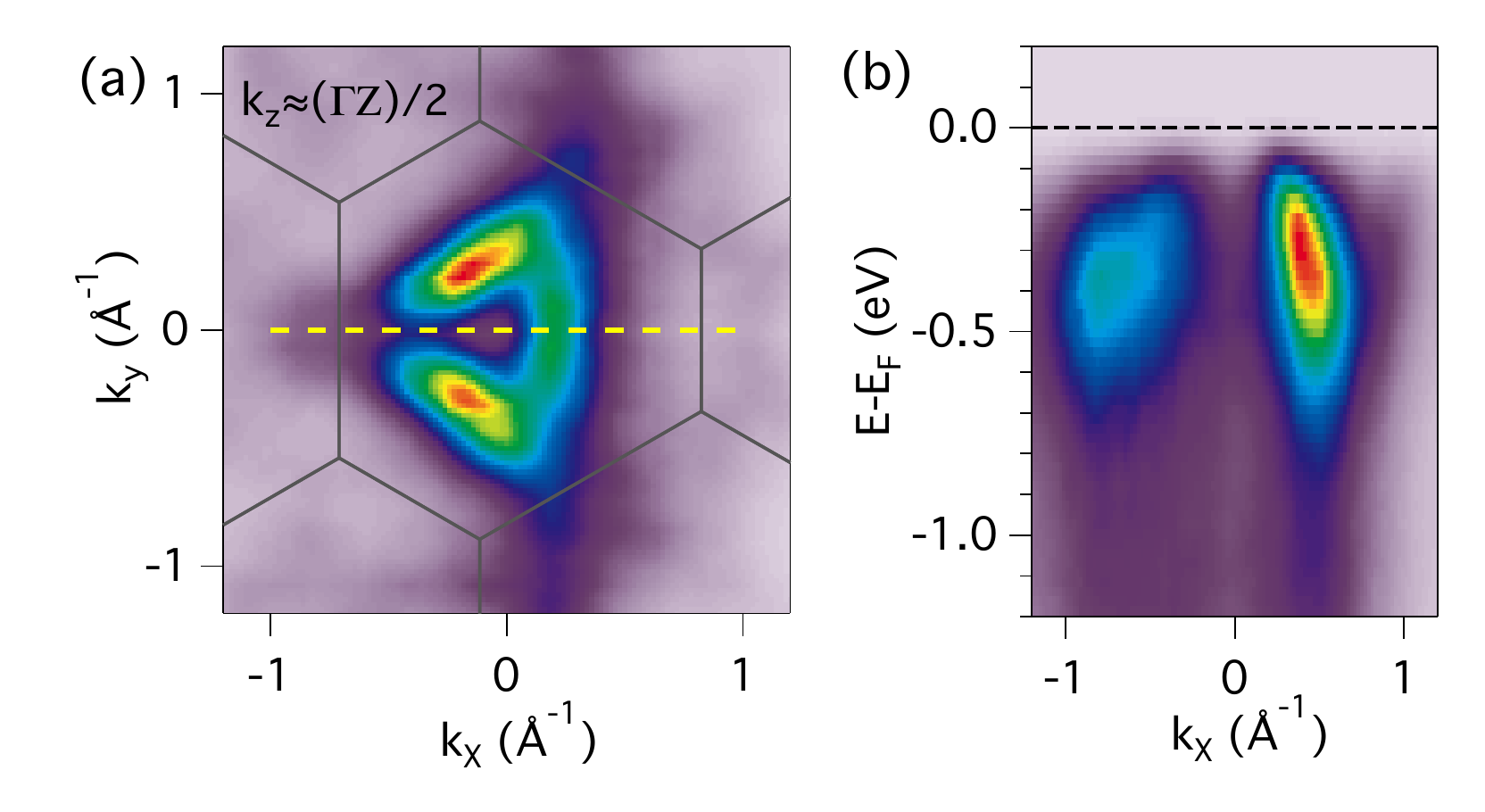}  
\end{center}
\caption{(a) Triangular feature measured at 180 eV showing three arc-like pieces converging at the corners of the triangle. (b) Hole-like energy dispersion cut through the triangular feature as indicated by the yellow dashed line in panel (a).
}
\label{Fig3}
\end{figure} 

The Fermi contour at $k_z$=$\Gamma$Z/2 is shown in Fig. 3(a). Data are taken at a lower photon energy (h$\nu$=180 eV) than that in Fig. 1. The overall contour resembles a triangular like shape with open tips, thus suggesting three arc-like features. A clear hole-like dispersion is visible in the energy distribution curve cut in Fig. 3(b) corresponding to the dashed line in panel (a).  The Fermi velocity is 1.63 eV-\AA. This FS feature is stable with small percentages of doping within the metallic phase, as indicated by data taken on (V$_{0.988}$Cr$_{0.012}$)$_2$O$_3$ and (V$_{0.955}$Ti$_{0.045}$)$_2$O$_3$ (not shown). The very existence of this triangular FS contour halfway between $\Gamma$ and Z strongly contradicts the DFT + DMFT calculation reported in [\onlinecite{poteryaev-07}]  in which only a single zone-centered \a1g\ electron FS is remaining while the top of the \epg\ states is pushed fully below \EF. Moreover the predicted depth of the \a1g\ electron band at $\Gamma$ is shallower than the experimental ARPES value of 0.4 eV \cite{suppl}.

\begin{figure}[b]
\leavevmode
\centering
\begin{center}
\includegraphics[width=8.4cm]{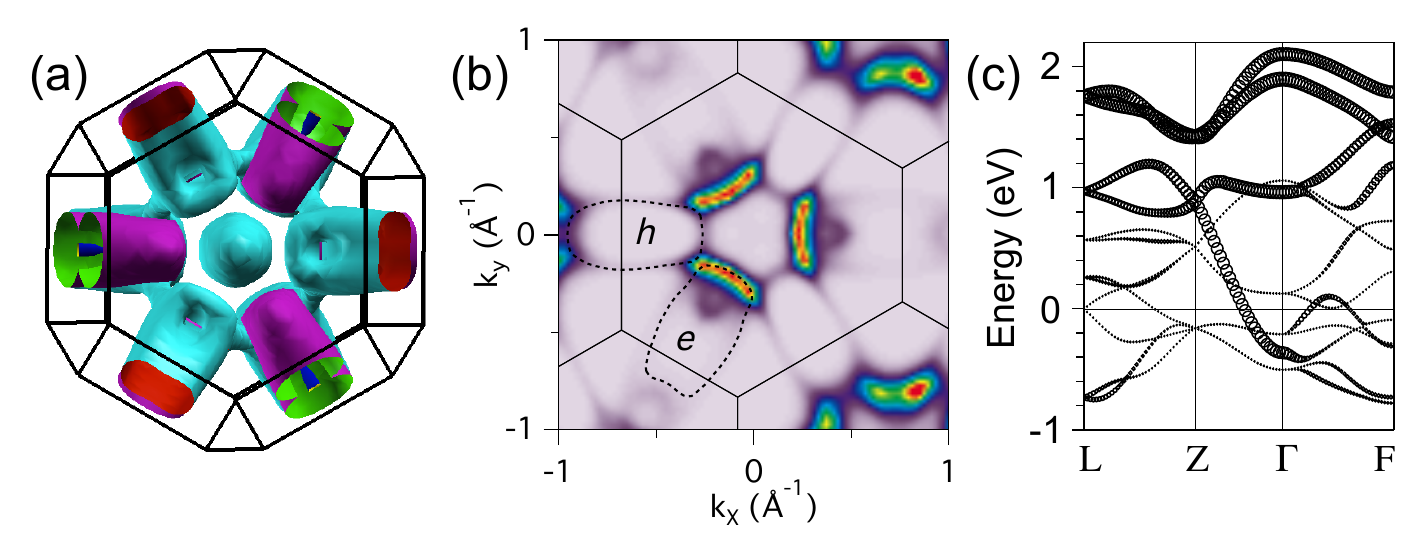}   
\end{center}
\caption{(Color Online)
(a) Top view of the theoretical Fermi surface of a non-magnetic DFT+$U$ (6.5 eV) calculation including both electron (blue) and hole (purple) sheets. (b) $k_z$=0.4 \invA\ image cut of Lorentzian \textbf{$k$}-broadened Fermi-edge contours. (c) V 3$d$ band structure for the same calculation with highlighting of the \a1g\ orbital character (fat line).
}
\label{Fig4}
\end{figure}

We have explored constant energy contours of various non-magnetic DFT+$U$ calculations, under the assumption that the DMFT energy renormalization, quasiparticle weight reduction and spectral weight redistribution, absent in DFT, does not radically change the shape of the QP band dispersions.
Fig. 4(a) and (b) show the 3D Fermi surface of a DFT+$U$ (6.5 eV) calculation \cite{wien2k} with constrained zero moment and a $k_z$=0.4 \invA\ (0.6$\cdot\Gamma Z$) ``spectral'' image cut of Lorentzian-broadened Fermi-energy contours. The calculation exhibits three-fold symmetric electron (e) and hole (h) sheets pointing upwards away from the center $\Gamma$-Z axis towards the square and hexagonal zone-boundary faces, respectively. The spectral weight of the inner edges of the electron sheets, containing vertical edges along the $c$-axis, are strongly enhanced with $k_z$-broadening \cite{suppl} relative to the highly $k_z$-dispersing parts of the electron or hole sheets.
These three-fold arcs give rise to the appearance of an open-tipped triangular Fermi surface that is very similar in size and shape to that of the ARPES measurement shown in Fig. 3(a). 
The increased $U$ value relative to that in earlier DMFT calculations\cite{poteryaev-07} produced the best agreement with the experimental results and is generally consistent \cite{suppl} with the values used in more recent DMFT calculations \cite{grieger-14, deng-14, leonov-15}

The band dispersion for the DFT+$U$ calculation is shown in Fig. 4(c).  The lowest of 4 \a1g\ bands crosses \EF\ approximately half-way along $\Gamma$-Z and the other 3 bands are unoccupied.  The zone-centered electron sheet corresponding to this partially occupied \a1g\ band is visible in the 3D Fermi surface in Fig. 4(a). The plot of band dispersions also clearly shows that \EF\ then cuts in the middle of the \epg\ bands.   It is very encouraging that two more recent $k$-resolved DFT + DMFT calculations \cite{grieger-14,deng-14} incorporating full charge self-consistency predict  \epg\ QP bands above and crossing \EF\ corresponding to less than half-filling of the \epg\ QP states \cite{suppl}. Our experimental result supports this scenario and does not support the earlier filled \epg\ QP band scenario \cite{poteryaev-07} in which the PM phase is at the threshold of the formation of an \a1g-\epg\ insulator gap at \EF. 

The mid-filling of the \epg\ QP states in the PM phase,  as opposed to being at the brink of an \a1g-\epg\  direct gap, challenges theory anew as to how the lattice constant or magnetic ordering perturbations of the PI and AFI phases are able to open an insulating gap.   A possible answer may lie with new charge self consistent but non-$k$-resolved GGA + DMFT calculations that find the MIT to be driven by a strong orbital selective renormalization featuring the \epg\ QP weight going to zero \cite{leonov-15}.  A calculation of $k$-resolved QP bands for comparison to our data would be required to more fully evaluate this line of study.  Recent hard x-ray photoemission and absorption experiments have provided additional MIT model constraints on the variation of $U$ between the metal and insulating phases \cite{fujiwara-11} and on the different variation of the orbital occupancies for the pressure-induced PI phase \cite{rodolakis-10}  as compared to the Cr-doping and temperature induced PI phases.  Thus the mystery of the MIT lives on. 

In conclusion we have reported experimental electron- and hole-like band dispersions in the metallic phase of \V2O3 at specific $k$-locations in the bulk Brillouin zone and with distinct Fermi surface topologies. We associate them with \a1g\ and roughly half filled \epg\ QP bands, thus supporting the scenario of partially filled \epg\ states, in disagreement with previously reported $k$-resolved DMFT calculations that described the metallic phase at the brink of a MIT driven by correlation-enhanced crystal field splitting. These results are a major new spectroscopic contribution to the long sought goal of a final and complete understanding of the (V$_{1-x}$Cr$_x$)$_2$O$_3$ MIT paradigm.

We gratefully acknowledge theoretical advice and stimulating discussions with Byung Il Min, Richard Martin and especially Frank Lechermann. This work was supported by the National Science Foundation grant number DMR-1410660. 
 J. W. A.  thanks the Advanced Light Source (ALS) for supporting a visit to the ALS during the time that this paper was written.
Theory calculations used resources of the National Energy Research Scientific Computing Center (NERSC). 
Both the ALS and NERSC user facilities are supported by the Office of Science of the U.S. Department of Energy under Contract No. DE-AC02-05CH11231.


\begin{thebibliography}{}
\bibitem{QPweights}
	As a rough illustration that considers only the \epg\ states which account for very nearly all 8 of the electrons per unit cell, i.e. neglects the small fraction in the \a1g\ QP bands, it is consistent that the \epg\ QP weight $Z$ is 0.2 and that the division of the $k$-integrated spectral weight below \EF\ is roughly equal between the QP bands and the lower Hubbard band.
\bibitem{kz}At normal emission ($k_{||}$=0) and lower photon energy $k_z$ in \AA\ is related to the photon energy in eV by $k_z$=0.512$\sqrt(h\nu-\phi+V_0)$, where $\phi$=4.5 eV is the work function of the analyzer and $V_0$=14 eV is the so-called inner potential.  At higher photon energy this basic relation has been modified to include quantitatively important effects of the photon momentum.
\end{thebibliography}
\end{document}


\DeclareGraphicsExtensions{.eps, .png}
\bibliographystyle{prsty}

\title{{\it Supplementary Material for}\\
``Fermi Surface of Metallic V$_2$O$_3$ from Angle-Resolved Photoemission:  \\
Mid-level Filling of $e_g^{\pi}$ Bands''}

\author{I. Lo Vecchio}
\affiliation{Materials Sciences Division, Lawrence Berkeley National Laboratory, Berkeley, California 94720, USA}

\author{J. D. Denlinger}
\affiliation{Advanced Light Source, Lawrence Berkeley National Laboratory, Berkeley, California 94720, USA}

\author{O. Krupin}
\affiliation{Advanced Light Source, Lawrence Berkeley National Laboratory, Berkeley, California 94720, USA}

\author{B. J. Kim}
\affiliation{Max-Planck-Institut fur Festk\"orperforschung, Heisenbergstrasse 1, D-70569 Stuttgart, Germany}

\author{P. A. Metcalf}
\affiliation{School of Materials Engineering, Purdue University, West Lafayette, Indiana 47907, USA}

\author{S. Lupi}
\affiliation{CNR-IOM and Dipartimento di Fisica, Universit\`a di Roma ``Sapienza,"  I-00185 Rome, Italy}

\author{J. W. Allen}
\affiliation{Randall Laboratory of Physics, University of Michigan, Ann Arbor, Michigan 48109, USA}

\author{A. Lanzara}
\affiliation{Materials Sciences Division, Lawrence Berkeley National Laboratory, Berkeley, California 94720, USA}
\affiliation{Department of Physics, University of California Berkeley, Berkeley, California 94720, USA}

\maketitle

\def\V2O3{V$_2$O$_3$}
\def\EF{$E_\textrm{F}$}
\def\vF{$v_\textrm{F}$}
\def\kF{$k_\textrm{F}$}
\def\invA{\AA$^{-1}$}
\def\etal{$et\ al.$}
\def\a1g{$a_{1g}$}
\def\epg{$e_g^{\pi}$}
\def\cred{\color{red}}
\def\cblue{\color{blue}}
\definecolor{dkgreen}{rgb}{0.2,0.7,0.2}
\def\cgreen{\color{dkgreen}}

\textbf{Contents} \\
1. $k$-resolved DMFT calculations \\
2. DFT calculations \\
3. $k_z$-broadening \\
4. Orbital occupations \\

\subsection{1. $k$-resolved DMFT calculations}

\begin{figure*}[t]
\leavevmode
\centering
\begin{center}
\includegraphics[width=16 cm]{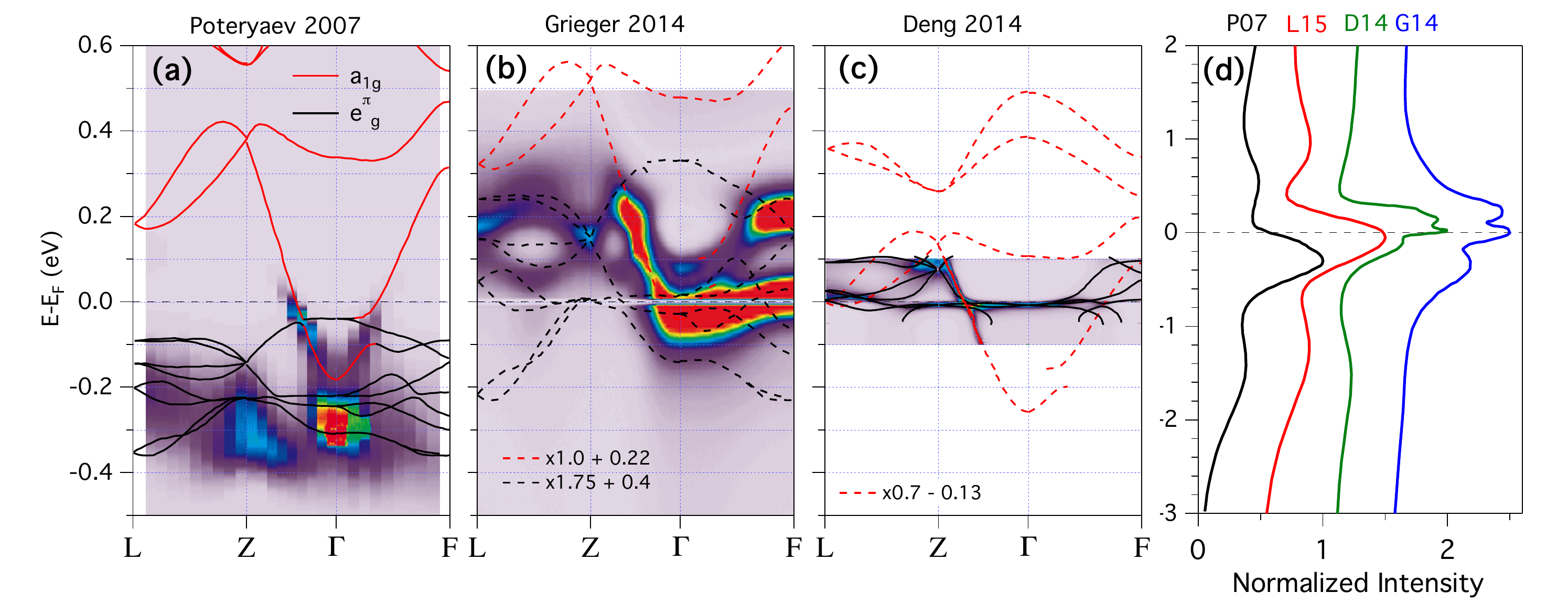}  
\end{center}
\caption{
Comparison of $k$-resolved DMFT spectral image plots along  LZ$\Gamma$F for \V2O3\ from (a) Poteryaev \etal\ \cite{Poteryaev07}, (b) Greiger \etal\ \cite{Grieger14}, and (c) Deng \etal\ \cite{Deng14}. Reported QP band energies for (a) and (c) are overplotted as solid lines, whereas dashed lines are scaled and shifted energies from (a).  (d) Comparison of the $k$-integrated $t_{2g}$ spectral weight profiles for the three calculations plus an additional DMFT calculation by Leonov \etal\ \cite{Leonov15}. 
}
\label{kdmft_compare}
\end{figure*} 

Three theoretical predictions of $k$-resolved band structures are currently found in the dynamical mean field theory (DMFT) literature of \V2O3 \cite{Poteryaev07,Grieger14,Deng14} and are plotted in Fig. \ref{kdmft_compare}(a-c).  The 2007 Poteryaev \etal\ study \cite{Poteryaev07} is a Hamiltonian-based DMFT calculation with a single-shot pass through the impurity solver step.  The 2014 Grieger and Lechermann calculation \cite{Grieger14}, with a focus on Cr-doping effects, explicitly highlights the improvement of full charge self-consistency through an outer loop that recalculates the DFT potentials for multiple DMFT passes.  The Deng \etal\ calculation \cite{Deng14}, found in the supplement of an optical study, is also charge self-consistent (although this is not explicitly stated).  
	A fourth 2015 Leonov \etal\ DMFT calculation \cite{Leonov15} that lacks $k$-resolved spectra, but also stresses full charge self-consistency, is included in the complementary $k$-integrated spectral weight profile comparison in Fig. \ref{kdmft_compare}(d).    The larger energy scale of $k$-integrated plot also shows spectral weight transfers to  higher and lower energies correspond to upper and lower Hubbards bands that are not discussed in this manuscript.

The three $k$-resolved DMFT images  were converted to grayscale and then replotted with a common color table and energy axis scale.  All were calculated along the same $L$-$Z$-$\Gamma$-$F$ high symmetry lines of the primitive rhombohedral Brillouin zone, where in the hexagonal BZ representation of Fig. 1,  $\Gamma$-$Z$ is along the $c$ axis, $L$ is the hexagonal face center, and $F$ is the $X$-point rectangular face center.  
To assist in the interpretation of the spectral image plots, reported quasiparticle (QP) pole energies from Poteryaev \etal\ and Deng \etal\ are overplotted as solid lines with \a1g\ (red) and \epg\  (black) color-coding.  Dashed-line bands represent scaled and shifted band energies from Poteryaev \etal.
The first general observation from the $k$-resolved and $k$-integrated plots is that there are significant differences in the centroid energy and bandwidth of the V 3$d$  states among all four theoretical predictions.

The  DMFT calculation by Poteryaev \etal\ in Fig. \ref{kdmft_compare}(a) exhibits (i) high spectral weight below \EF\ coming from the \epg\ bands (black), (ii) an electronlike dispersion at the $\Gamma$ point with a band minimum of $\approx$ -0.18 eV coming from the \a1g\ states (red), and (iii) a rapid energy broadening above \EF\ limiting the observation of the \a1g\ bands to less than 0.1 eV above \EF.  From the comparison of the QP pole energies to DFT theory, Poteryaev \etal\ note bandwidth energy renormalizations of 2.5$\times$ and 5$\times$ for the \a1g\ and \epg\ states, respectively. 
The entire \epg\ 8-band manifold is observed to be entirely below \EF, resulting in a Fermi surface arising solely from the \a1g\ QP electron dispersion.  
 This relative separation of downwards shifted of \epg\ states and upwards-shifted \a1g\ states, represents the scenario of strong correlation-enhanced trigonal crystal field splitting resulting in enhanced polarization of the \epg\ and \a1g\ orbital occupations.  In this scenario, common with other DMFT calculations \cite{Held01,Keller04,Laad03,Laad06,Hansmann13}, the PM phase is not far from a metal-insulator transition (MIT) to a state with a complete \epg\ - \a1g\ gap.  Also as discussed in the main text, the full occupation of the \epg\ states (representing 16 electron states total) seemingly violates, by a factor of 2, the near $d^2$ electron configuration of the four V$^{3+}$ atoms per unit cell (8 electrons total).  However the much reduced $Z$$\approx$0.2 quasiparticle weighting of the bands  allows this scenario (including spectral weight transfers to higher energies) without violating the Luttinger count, as explained explicitly in a footnote of the main text.

In contrast, the DMFT calculation by Grieger \etal\  in Fig. \ref{kdmft_compare}(b) shows (i) high spectral weight $above$ \EF\ with unoccupied band dispersions visible up to +0.3 eV,  and (ii) a high intensity $\Gamma$ point electron dispersion with shallower $<$0.1 eV minimum energy with somewhat flat dispersion near \EF\ along $\Gamma$-$F$.  
Grieger \etal\ did not provide QP pole energies, so we make use of overplotting the Poteryaev \etal\ QP energies with separate scaling and energy shifts to assist in discussing the differences. 
 First we shift the \a1g\ bands to higher energy to match the -50 meV electron band minimum, and observe that the shifted \a1g\ bands along $L$-$Z$ are too high in energy to account for the $<$0.3 eV band dispersions which then must be of \epg\ origin.  Also the \a1g\ bands cannot account for the high intensity flat bands close to \EF\ along $\Gamma$-$F$.  We then shift the Poteryaev \etal\ QP \epg\ bands to match the +0.15 eV enhanced intensity \epg\ degenerate point at $Z$, and expand the \epg\ bandwidth by 1.75$\times$ to match the high intensity \EF\ flat bands along $\Gamma$-$F$. The larger \epg\ band width indicates a weaker energy renormalization ($\approx3\times$ relative to DFT) compared to that of Poteryaev \etal.  The positive energy shift of the \epg\ bands relative to Poteryaev \etal\ indicates a Fermi-level that lies near the middle range of the \epg\ QP band manifold.

The third DMFT calculation by Deng \etal\  in Fig. \ref{kdmft_compare}(c) was plotted with too narrow an energy range to allow direct inspection of relative spectral weights above or below \EF, or the energy depth of the $\Gamma$-point electron dispersion.  Deng \etal\ did plot QP pole energies onto their $k$-resolved spectral image, which we enhance with black solid lines, allowing recognition of the characteristic \epg\ $Z$-point degeneracy energies of +0.08 eV and a narrow cluster of \epg\ bands just below \EF\ similar to Grieger \etal.  These flat \epg\ states produce a  peak at \EF\ in the $k$-integrated spectral weight profiles in Fig. \ref{kdmft_compare}(d) for both calculations.
 However the overall \epg\ band width is approximately 2$\times$ narrower than that of Grieger \etal\ indicating an energy renormalization of the \epg\ bands similar to that of Poteryaev \etal. 

We also identify a portion of the highly dispersive \a1g\ electron band in red along $\Gamma$-$Z$, and  extrapolate this band (dashed red curve) based on the Poteryeav \a1g\ dispersion by matching the slope and \kF-crossing point along $\Gamma$-$Z$.  The resulting 0.7$\times$ smaller bandwidth and negative energy shift suggests an \a1g\ band minimum as deep as 0. 25 eV below \EF,  and possibly deeper than the bottom of the \epg\ QP manifold.  In comparison, the experimental ARPES $\Gamma$-point electron pocket depth of $\approx$0.4 eV and hole band dispersion down to 0.5 eV are deeper than any of the these $k$-resolved DMFT calculations.

The $k$-integrated spectral weight profile from the 2015 charge self-consistent DMFT calculation of Leonov \etal, plotted in Fig. \ref{kdmft_compare}(d), lacks the characteristic narrow peak at \EF\ that is found in the profiles of Grieger \etal\ and Deng \etal\ (arising from flat \epg\ bands), and has its energy centroid slightly below \EF\ yet noticeably higher in energy than occurs in Poteryeav \etal.   
Hence the spectral weight profile of Lenov \etal\ is also suggestive of the Fermi level cutting in the middle of the \epg\ band manifold, although slightly higher than in Grieger \etal\ and Deng \etal.

The lack of experimental ARPES evidence for the sharp DOS peak at \EF\ predicted by two of the DMFT calculations and the deeper energy dispersions of the ARPES \epg\ hole and \a1g\ electron bands indicate the importance of further theoretical refinement.  The newer charge self-consistent DMFT calculations provide a common theme of the Fermi level residing in the middle of the \epg\ QP bands, in contrast to the earlier Poteryaev \etal\ calculation, and qualitatively consistent with our experimental conclusion based on matching the ARPES Fermi surface to non-magnetic DFT + $U$ calculations.     This calls into question the scenario of the correlation-enhanced trigonal crystal field splitting pushing the PM phase close to the threshold of an \epg\ - \a1g\ insulating gap, i.e. that the orbital polarization effect is significantly weaker.  Indeed Grieger \etal\ calculate a DMFT orbital occupation ratio for pure \V2O3\ as low as \epg:\a1g= 2.4:1, significantly smaller than that of Poteryaev \etal\ (3.8:1) and even smaller than the polarized-XAS result of 3:1 \cite{Park00}. A more general summary and discussion of \epg:\a1g\ ratios for various calculations is given in a separate section below.  
    
We close this discussion by noting that increased values of $U$ relative to those of earlier DMFT calculations such as Poteryaev \etal\ ($U$=4.2 eV) are a common feature of the recent charge self-consistent DMFT calculations which employ $U$=5 eV by Grieger \etal\ and Leonov \etal,  and $U$$>$6 eV by Deng \etal\ for the PM phase.

\section{ 2. DFT Calculations}

\begin{figure}[b]
\leavevmode
\centering
\begin{center}
\includegraphics[width=14.5 cm]{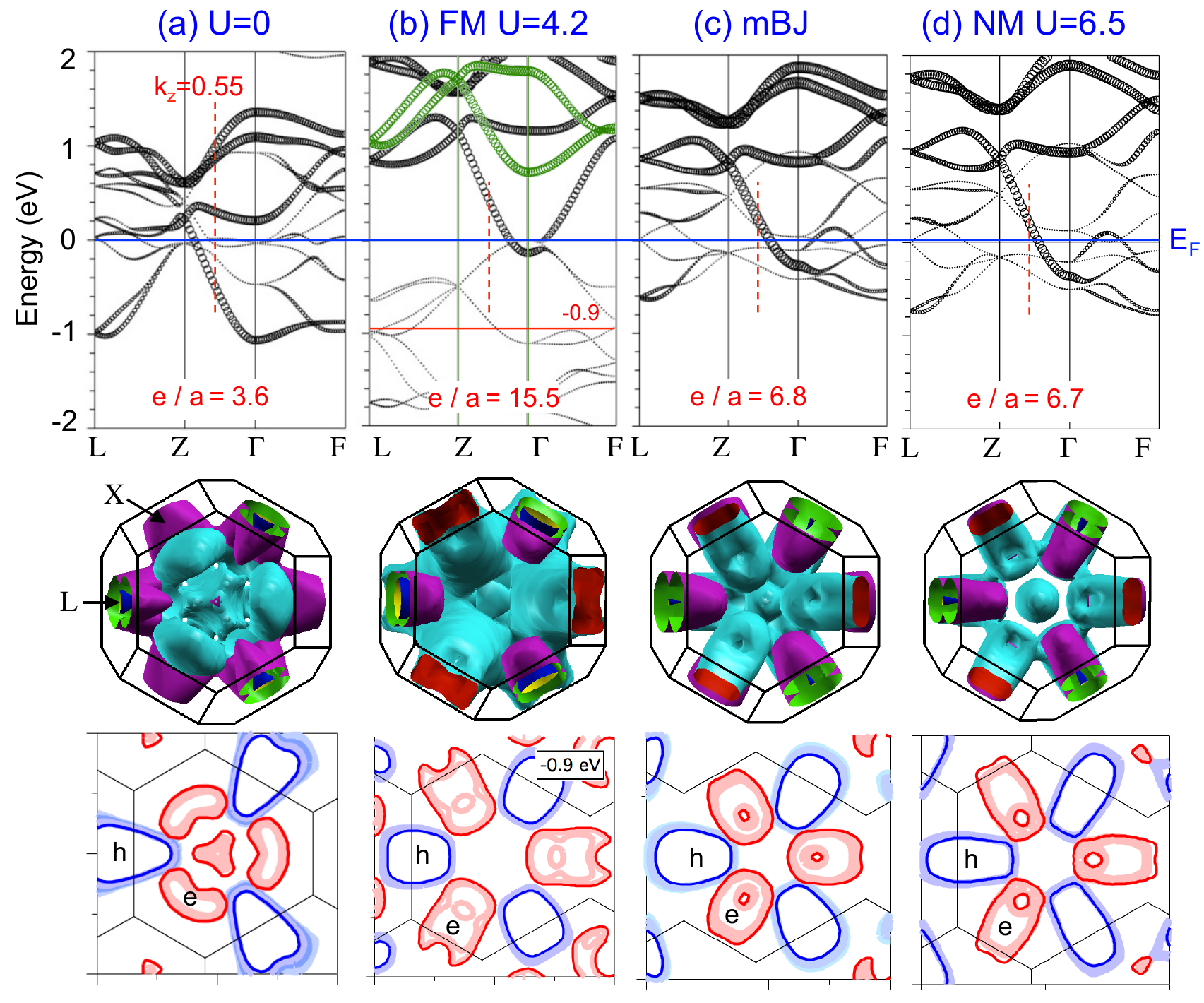} 
\end{center}
\caption{
Comparison of various DFT calculations including  (a) uncorrelated GGA ($U$=0), (b) spin-polarized ferromagnetic GGA + $U$ (4.2 eV), (c) GGA + mBJ and (d) non-magnetic GGA + $U$ (6.5 eV).  
The comparison includes (top) the V 3$d$ band structure with fat line highlighting the \a1g\ orbital character,  (middle) Top view of the 3D Fermi surface, and  (bottom) FS contours at $k_z$=0.3-0.37 \invA (0.45-0.55 $\times\Gamma$-$Z$) with lighter shading represents deeper binding energies indicating electronlike (red) or holelike (blue) band dispersion. 
The occupied orbital character ratio e$^\pi_g$ : \a1g\ is also listed ($e/a$) for each calculation. 
}
\label{dft_compare}
\end{figure} 

In this supplemental section and Fig. \ref{dft_compare} we provide a brief comparison of various density functional (DFT) and correlated DFT + $U$ calculations that were investigated in this study with the primary goal of exploring the various QP constant energy contours one might expect from the various DMFT theoretical scenarios of different \a1g-\epg\ orbital polarization energy separations.  We stress that the presented DFT calculations are just a tool for investigating QP energy contours, especially useful for the non-high symmetry $k$-space planes suggested to be of interest by ARPES and not currently in the published $k$-resolved DMFT calculations.  These calculations are not expected to capture the physics of energy renormalization, or reduced QP weighting of bands with corresponding spectral weight transfers to higher energy scale lower and upper Hubbard bands. The DFT calculations were performed using the full-potential linearized augmented plane-wave band method as implemented in the Wien2K package \cite{wien2k}.  The PBE-GGA (Perdew-Burke-Ernzerhof 96) exchange-correlation potential was used along with the self-energy correction (SIC) method for the DFT + $U$ calculations.  Additional corrections are discussed below.

{\bf GGA ($U$=0)}.  Fig. \ref{dft_compare}(a) shows a standard uncorrelated GGA reference calculation whose band structure well reproduces published results \cite{Mattheiss94,Dasgupta09} and whose 3D FS topology (middle row) contains two complex electron (blue) and hole (violet) sheets similar to those of a recently published FS plot \cite{Grieger15}.  A constant $k_z$=0.3 \invA\ (0.45$\times\Gamma$-$Z$) cut of this GGA ($U$=0) Fermi surface is shown in the bottom row.  It contains multiple contours including a triangularlike $closed$ contour with decent agreement in size to ARPES, but with  electronlike band dispersion, as indicated by the deeper binding energy  shading interior to the \EF\ contour, that is opposite to the experimental ARPES result.  That contour originates from the tip of a large zone-centered ellipsoidal FS formed by the large electronlike \a1g\ dispersion along $\Gamma$-$Z$ with the \kF-crossing point very close to $Z$ (top row panel).  Thus we conclude that the uncorrelated reference calculation cannot explain the open-tipped holelike triangular feature from ARPES.
Also notable for the $U$=0 calculation is the existence of flat \epg\ bands near \EF\ along $Z$-$\Gamma$-$F$ similar to the results in the Grieger \etal\ and Deng \etal\ DMFT calculations.

{\bf FM GGA + $U$}.  After the reference calculation, we first tried a ferromagnetic GGA + $U$ calculation, shown in Fig. \ref{dft_compare}(b), that mimics the strong \a1g-\epg\ orbital separation of the Poteryaev \etal\ DMFT calculation.    The FM calculation reproduces NMTO spin-polarized results by Elfimov \etal\ \cite{Elfimov03} where only the spin-up states close to \EF\ are retained.  The spin down components of both \epg\ and \a1g\ states are pushed higher above \EF.  The correlation energy of $U$=4.2 eV, similar to that used in DMFT calculations, results in the near separation of the spin-up \a1g\ and \epg\ states, poising it close to an MIT.  The FS consists of a single zone-centered \a1g\ electron pocket due to the \epg\ states being entirely below \EF.  No correspondence to the ARPES holelike triangular contours is found near \EF\ (not shown), either by tuning the contour energy down into the top of the \epg\ states, or by decreasing the $U$ value to bring the \epg\ states up to crossing \EF.    
However the deep -0.9 eV contour in the middle of the \epg\ manifold is shown in the lower two panels to illustrate its characteristics that are similar to the Fermi-edge contours of the next two presented DFT calculations.  
	We also note that the nearly complete filling of \epg\ bands simultaneous with a small \a1g\ occupation can satisfy the $\sim$$d^2$ electron configuration in this FM calculation by virtue of the spin-polarization, i.e. one electron in each of 8 \epg\ bands, whereas reduced QP weighting allows such a scenario of nearly filled spin-degenerate \epg\ bands  in the Poteryaev \etal\  and other DMFT calculations.  

{\bf GGA + mBJ}.  After the failure of the FM DFT + $U$ calculations to explain the ARPES, a non-magnetic GGA calculation employing the modified Becke-Johnson (mBJ) correction \cite{Tran09} was next attempted.  The mBJ method corresponds to an orbital-independent semi-local exchange-correlation potential  that mimics the behavior of orbital-dependent potentials and has been shown to provide semiconductor band gap corrections in good agreement with the more computation demanding many-body GW calculation.  
With reference to the $U$=0 calculation, the mBJ calculation in Fig. \ref{dft_compare}(c)  shows a significant modification to the electron FS sheet and only minor modification to the hole FS sheet. The zone-centered electron pocket is observed to shrink in size and become absent in the $k_z$=0.37 \invA\ FS contour plot.  In contrast the three rounded GGA nodules are observed to expand in size to connect to the second BZ through the F-point face of the bulk BZ.  The edges of these electron sheets diagonally pointing to the zone center, form holelike arcs relative to the central  $c$ axis.  The three-fold symmetry of these arcs gives the appearance of an open-ended triangular-shaped holelike FS contour that is midway along $\Gamma$-$Z$ and similar to the ARPES result.  
    The mBJ calculation notably has its Fermi level in the middle of the \epg\ bands in contrast to the FM DFT + $U$ calculation and it gives the appearance of some energy renormalization of the occupied \epg\ states.  This is in fact just an ``unhybridization'' effect of the \a1g\ and \epg\  orbitals \cite{Poteryaev07}, where the smaller isolated bandwidth of the \epg\ states is being restored as the overlap to the upwards shifting \a1g\ states becomes less.

{\bf NM GGA + $U$}.  Motivated by the success of the mBJ calculation, we finally investigated non-magnetic GGA + $U$ calculations in which the spin-polarization is constrained to be zero.  We empirically found  that the effect of $U$ in producing orbital polarization changes was much reduced compared to that of the FM GGA + $U$ calculation, and had to be increased to higher values. We discovered that larger values of $U$ between 6 eV and 7 eV produced results strikingly similar to those of the nonmagnetic GGA + mBJ calculation.  Increasing values of $U$ shrink the size of the zone center \a1g\ electron sheet and increase the separation of the triangularlike edges of the $L$-point electron sheets.  The final value of 6.5 eV shown in Fig. \ref{dft_compare}(d) produces the best match to both the size of the ARPES triangular FS and to the mBJ band energies. The Fermi level that produces the triangular FS  is found to be located in the middle of the \epg\ bands, also similar to what is found in the mBJ calculation.  The large value of $U$ required to produce separation of the \epg\ and \a1g\ bands {as} compared to that of the FM DFT + $U$ calculation likely arises from the spin-degeneracy and the constraint that \EF\ lies in the middle of the \epg\ states.  
As noted already above, the recent charge self-consistent DMFT calculations, similarly exhibiting Fermi level energies in the middle of the \epg\ bands, also employ  larger values of $U$ as compared to earlier DMFT calculations such as that of Poteryaev \etal\ ($U$=4.2 eV) with values for the PM phase as high as $U$=6.4 eV at 200 K in a $T$-$U$ phase diagram constructed by Deng \etal. 

The empirical observation of the mBJ calculation to mimic so cleanly the effect of $d$ correlations without specification of a $U$ parameter is also not readily found in the literature.

\subsection{3. $k_z$-broadening}

\begin{figure}[b]
\leavevmode
\centering
\begin{center}
\includegraphics[width=14.5 cm]{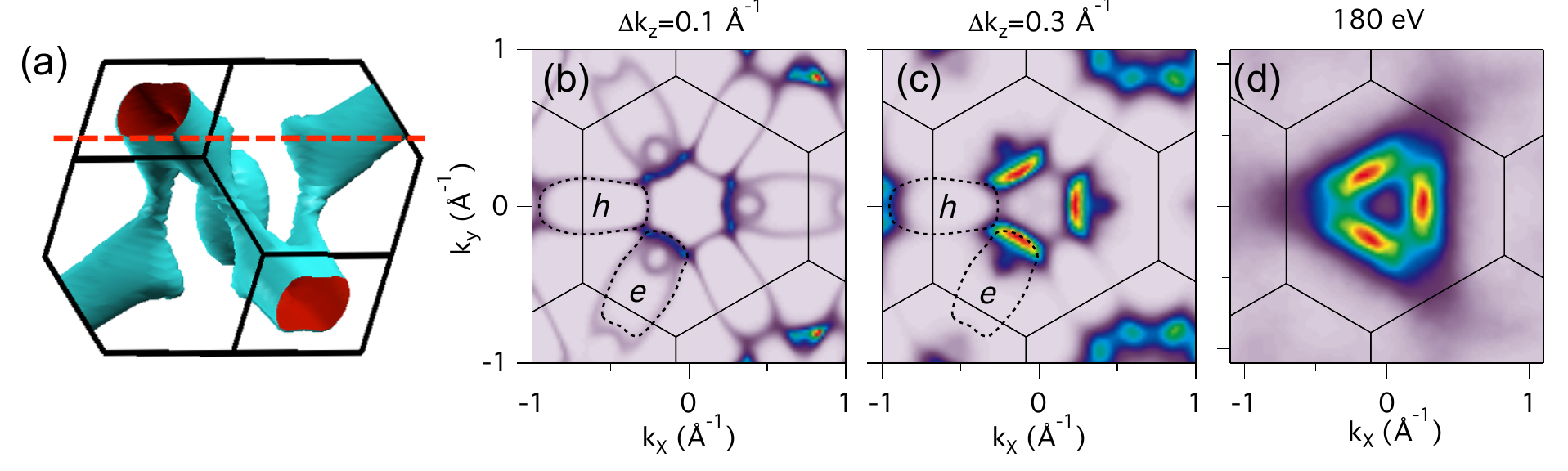}  
\end{center}
\caption{
Comparison of the GGA + $U$=6.5 eV calculated Fermi surface to ARPES.  (a) Side view of the electron topology FS sheet. (b,c) Pseudo-spectral intensity maps at $k_z$=0.6 $\Gamma$-$Z$ (0.4 \invA, red line in (a)) with Lorentzian-weighted $k_z$-averaging widths of (b) 0.1 \invA\ and (c) 0.3 \invA. (d) Three-fold symmetrization of the 180 eV ARPES FS map from Fig. 3(a). 
}
\label{kzbroaden}
\end{figure}

The good qualitative agreement between the ARPES Fermi surface in Fig. 3(a) and the DFT + $U$ (6.5 eV) Fermi surface presented in Fig. 4(b) results from two key factors of (i) the Fermi  level cutting through the middle of the \epg\ manifold, and (ii) the global FS topology that allows the edges of the $L$-point hole sheets to become significantly enhanced from the effect of $k_z$-broadening, which results from the small inelastic mean free path of the photoelectrons and the breaking of translation symmetry by the surface.
Figure \ref{kzbroaden}(a)  shows the side view the electron sheet of the Fermi surface. Midway between the $\Gamma$ and $Z$ planes as indicated by a red line, nearly vertical edges are observed in the $L$-point electron sheets.  The FS contours are repeated in Fig. \ref{kzbroaden}(b) for $k_z$=0.4 cutting midway between the $\Gamma$ and $Z$ planes showing both electron and hole sheet contours. 
Extraction of a pseudo-spectral image cut from the 3D volume of Fermi surface $k$-points  is done with a Lorentzian point spread function with width ($\Delta$$k_z$) that also provides the effect of averaging different values of $k_z$ perpendicular to the surface.  
Two different Lorentzian broadenings of 0.1 \invA\  and 0.3 \invA, corresponding to approximate mean free paths (1/$\Delta$$k_z$) of 10 \AA\ and 3.3 \AA, respectively, are shown  in Fig. \ref{kzbroaden} whereas the intermediate value of 0.2 \invA\  is used for Fig. 4(b).  
The result for increasing $k_z$-broadening is a dramatic spectral enhancement of the vertical edges of the X-centered electron sheets and weakening  of spectral weight where there is greater $k_z$-dispersion including the entire $L$-centered hole sheets.  Finally we make comparison in Fig. \ref{kzbroaden}(d) to the experimental ARPES 180 eV Fermi surface from Fig. 3(a) with a  three-fold symmetrization to remove effects of the photon-polarization and experimental geometry of the incident light.  Additional experimental thermal broadening comes from the 200 K sample temperature required to stay above the antiferromagnetic insulating phase below $\sim$165K.

\subsection{4. Orbital occupations }

The  ARPES presented in this paper provides new experimental quantitative checkpoints on future theoretical calculations for the depth of the $\Gamma$-point \a1g\ electron dispersion and the location of \EF\ within the \epg\ manifold. 
Previously, linearly-polarized x-ray absorption spectroscopy (XAS) measurements by Park \etal\ \cite{Park00} provided an experimental quantitative benchmark for the orbital occupation ratio ($e/a$ = \epg/\a1g\ = 3:1) for the metallic phase of \V2O3.  
It is interesting to understand where the new ARPES and various $k$-resolved DMFT calculations stand in relation to the XAS value.  However, the orbital occupations are not directly accessible from the ARPES measurements and  can only be deduced from theory calculations with which there is spectral agreement.  
Hence we provide in Table \ref{table1} a summary comparison of such orbital occupation quantities for the various DFT calculations presented in Section 2 and the four literature DMFT calculations in Section 1, to allow discussion of the differences and origins.

We first discuss two idealized cases referring to the GGA ($U$=0) calculation shown in Fig. \ref{dft_compare},  where two different \a1g\ band filling scenarios are visualized along different $k$ directions.
Along $L$-$Z$ well-defined bonding/anti-bonding gaps between both \a1g\ and \epg\ bands are present, with the larger \a1g splitting completely encompassing the narrower \epg\ states.   The half-filling of the \a1g\ states (2 of 4 bands)  
corresponds to the Castellani \cite{Castellani78} spin=1/2 model. 
The total 4 of 8 occupied \a1g\ electrons implies 4 of 16 occupied  \epg\ electrons for the $d^2$ configuration of the 4 V ions per unit cell. This is consistent with $\approx$2 of 8 \epg\ bands just below \EF\ along $L$-$Z$ in the GGA calculation. Thus the \epg\ and \a1g\ occupations are equal and there is no net orbital polarization ($e/a$=1:1).

Next we observe that along $\Gamma$-$Z$-$F$ in the same GGA ($U$=0) calculation,  only 1 of 4 bands with \a1g\ character appear below \EF. This quarter-filling of \a1g\ states (2 of 8 electrons) leaves 6 of 16 occupied \epg\ electrons (3/8 band filling)  and an orbital polarization ratio of $e/a$ =3:1. This coincidently equals the polarized-XAS result of Park \etal\ \cite{Park00} for the PM phase of \V2O3. Similar orbital occupation values of $(e,a)$=($\approx$0.5,$\approx$1.5) can be found in various $U$=0  DFT reference calculations in the literature \cite{Keller04,Poteryaev07,Grieger14}. This perplexingly suggests that any DFT + $U$ enhanced orbital polarization beyond the uncorrelated DFT calculation will naturally result in $e/a$ values larger than the polarized-XAS benchmark.

Indeed this is what is observed in the DFT  calculations shown in the middle block of Table \ref{table1}.
The Wien2k orbital occupations are computed within muffin-tin spheres whose results are sensitive to the muffin-tin radii. Hence the values have been scaled in Table \ref{table1} to give the $d^2$ occupation, $(e+a)$=2.0. 
The GGA ($U$=0) $e/a$ ratio of 3.6 is a slightly larger than the 1/4-filled \a1g\ scenario.  The nonmagnetic GGA + $U$ (6.5 eV) and GGA + mBJ give similar numeric values and correspond closely to a 1 of 8 \a1g\ electron occupation, leaving 7 of 16 occupied \epg\ electrons, and an orbital polarization ratio of $e/a$=7:1. Finally the ferromagnetic GGA + $U$ (4.2 eV) calculation with nearly separated \a1g\ and \epg\ states is very similar to an idealized scenario of only 0.5 of 8 \a1g\ and 7.5 of 16 \epg\ electron occupation, resulting in a very large $e/a$ orbital polarization ratio of $e/a$=15:1.
The basic observation is that all these orbital polarization values are much greater than the XAS result of 3:1, including the GGA + $U$ (6.5 eV) and GGA + mBJ calculations that we claim give good agreement to the ARPES Fermi surface topology.

To better understand the factors involved in this discrepancy, we next compare occupation values reported for the DMFT calculations of Section 1, listed in the third block of Table \ref{table1}.  An additional column lists the orbital-dependent reduced QP spectral weight ($Z$) values.   Poteryaev \etal\ actually report two different analyses of the orbital occupations.  The first set of $(a,e)$=(0.2,1.8)  (from Table I of Ref. \cite{Poteryaev07}) gives a very large ratio $e/a$=9 that is consistent with the near separation of \a1g\ and \epg\ states and the FM-GGA + $U$ calculation.  The second set of reported values $(a,e)$=(0.45,1.7)   results from a partial-wave analysis of localized V 3$d$ core electron occupation that is argued to be more appropriate for comparison to the XAS experiment (which creates a V 2$p$ core-hole hole in the final state).  It appears to significantly enhance the relative \a1g\ occupation and give a much reduced ratio of $e/a$=3.8. This represents one possible physics origin for the large DFT + $U$ $e/a$ values.

However, the newer self-consistent DMFT calculations with different mid-filling of the \epg\ QP bands, but without any stated special consideration of localized electron occupation, produce occupation values that are very similar to the uncorrelated GGA ($U$=0) result and thus $e/a$ ratios similar to those of the XAS result. Here we 
also note that the DMFT reduced QP weighting of the occupied valence bands in column $Z$ of Table \ref{table1}  has an orbital dependence.  The larger incoherency of the \epg\ states compared to  \a1g\ states, which results in different spectral weight transfers for the two orbitals to higher and lower energy Hubbard bands, may also play an important role in a general correction to the DFT-calculated orbital occupation values. 

In conclusion, there are multiple reasons why the numeric orbital occupation ratio of the DFT + $U$ calculation that exhibits agreement to the ARPES Fermi surface may not be comparable to the polarized-XAS benchmark of $e/a$=3:1, including (i) different DFT code methods for evaluating orbital characters, (ii) physics of the XAS process, and (iii) reduced QP weighting and spectral weight transfers not included in the DFT calculations.

{\setlength{\tabcolsep}{0.7em}
\begin{table}
\caption{Orbital occupation summary (per V atom) for various models and calculations.}
\begin{tabular}{ l cc cc c c l}

 \toprule
  \textbf{Model} & $U$ (eV) & $Z$(\a1g, \epg) & \a1g & \epg  & \epg/\a1g  \\ 
 \toprule
      Castellani \cite{Castellani78}  & 0 & 1 & 1.0 & 1.0  & 1 \\
      Park, XAS \cite{Park00}  & 0 & 1 & 0.5 & 1.5  & 3 \\ 
\colrule
     GGA  & 0 & 1 & 0.431 & 1.569  & 3.6 \\ 
     NM-GGA+$U$ & 6.5 & 1 & 0.259 & 1.741   & 6.7 \\  
     NM-GGA+mBJ & - & 1 & 0.256 & 1.744  & 6.8 \\  
     FM-GGA+$U$ ($\uparrow$)$^a$ & 4.2 & 1 & 0.121 & 1.879   & 15.5  \\ 
\colrule
     DMFT, Poteryaev \cite{Poteryaev07} & 4.2 & 0.4, 0.2 & 0.2 & 1.8  & 9.0  \\ 
     DMFT, Poteryaev \cite{Poteryaev07}$^b$ & 4.2 & 0.4, 0.2 & 0.45 & 1.70  & 3.8 \\ 
     DMFT, Grieger \cite{Grieger14} & 5 & 0.34, 0.23 & 0.59 & 1.41  & 2.4 \\  
     DMFT, Deng \cite{Deng14} & 6 & ?  & 0.5 & 1.6 & 3.2  \\
     DMFT, Leonov \cite{Leonov15} & 5 & 0.145, 0.125$^c$  & 0.44 & 1.56 & 3.5 \\
\botrule
\end{tabular}
\\{$^{\rm a}$Spin-down bands completely above \EF.}
\\[-6pt]{$^{\rm b}$Partial-wave analysis of local V 3$d$ core electrons.}
\\[-6pt]{$^{\rm c}$$Z$ values at the PM phase total energy minimum cell volume.}
\label{table1}
\end{table}
}
